\documentclass[aps,prx,groupedaddress,twocolumn,amsmath,ambsymb]{revtex4-2}

\usepackage{hyperref} \hypersetup{colorlinks=true, citecolor = blue, linkcolor=teal, filecolor=magenta, urlcolor=cyan}

\usepackage{graphicx}
\usepackage{color,colortbl}
\usepackage{xcolor}
\usepackage{soul}

\newcommand{\CstTop}{C^*_{5\%}}
\newcommand{\gammai}{\gamma_0}
\newcommand{\gammac}{\gamma_{\rm c}}
\newcommand{\gammamax}{\gamma_{\rm max}}

\newcommand{\ly}{l_{\rm y}}
\newcommand{\lw}{l_{\rm w}}

\newcommand{\siginit}{\sigma_{\rm init}}

\newcommand{\vecv}[1]{\mathbf{{#1}}}
\newcommand{\tens}[1]{\mathbf{{#1}}}

\begin{document}

\title{Mechanism of delayed elastic state absorption versus fatigue failure in cyclic shear of amorphous materials}

\author{Samuel B. Walker and Suzanne M. Fielding}%
\affiliation{Department of Physics, Durham University, Science Laboratories, South Road, Durham DH1 3LE, United Kingdom.}%

\date{\today}

\begin{abstract}
We study cyclic shear within a lattice elastoplastic model of amorphous materials. In numerical simulations, the model displays a threshold shear amplitude $\gammac$, which increases with increasing sample annealing prior to shear.  For imposed amplitudes $\gammai\lesssim\gammac$, it shows slow progression to an absorbing elastic state. For $\gammai\gtrsim \gammac$, it shows slow fatigue leading to delayed material failure. Our key contribution is to elucidate a mechanistic understanding of this behaviour. We show that progression to an absorbing elastic state maps to the physics of single element diffusion on a finite interval with two absorbing boundaries. This enables us to derive analytical expressions for the decay rate of the fraction of active elements as a function of cycle number,  and the survival distribution across samples, in excellent agreement with the numerical simulations. The relevant diffusing variable is that of local elemental strain, which takes two jumps stroboscopically in each cycle. Once the value at any lattice site diffuses below its local yielding threshold, the site's element behaves elastically in subsequent cycles. This leads to a gradual accumulation of strain values just below threshold, eventually for all elements when $\gammai<\gammac$. For $\gammai\lesssim \gammac$ this diffusive behaviour is preceded by many cycles in which cooperativity  between elements via spatial stress propagation is also important. For $\gammai > \gammac$, stress propagation dominates and precipitates material failure. For $\gammai\gtrsim \gammac$, the same slow accumulation of elements just below their local threshold that led eventually to an elastic state for $\gamma_0\lesssim\gamma_{\rm c}$ also accounts for the slow fatigue that brings the system to the brink of failure. We evidence this via simulations in which the strain amplitude artificially exceeds $\gamma_0 $ in each half cycle, before reverting to the system's actual state at $\pm\gamma_0$ and resuming the next half cycle as normal. During fatigue, the stress-strain slope $d\sigma/d\gamma$  evolves towards unity at $|\gamma|=\gamma_0^-$, signifying perfect elasticity within cycle, yet towards zero at $|\gamma|=\gamma_0^+$, precipitating mechanical instability and material failure.
\end{abstract}
\maketitle

\section{Introduction}
\label{sec:intro}

Amorphous materials~\cite{berthier2011theoretical,bonn2017yield,Nicolas2015} include soft solids such as colloids, emulsions, gels, granular materials, and biological tissues, as well as hard materials such as metallic glasses. Unlike conventional crystalline solids, these materials lack any long ranged ordering in the arrangement of their constituent substructures (colloidal particles, emulsion droplets, sand grains, biological cells, etc.). This makes understanding their mechanical properties a major scientific challenge. Typically, they behave elastically at small imposed deformations (strains) or loads (stresses), then fail in either a ductile or brittle way at larger strains or stresses~\cite{divoux2024ductile}. Considerable effort has been devoted to understanding this yielding transition as it manifests following the switch-on of a constant shear rate or stress in a previously unsheared material. In many practical applications, however, materials are subject to a cyclically repeating deformation or load. Cyclic shear is also important fundamentally: in affording the opportunity to understand the yielding transition in the context of a sustained, repeating shear, and in identifying the key fingerprints of a material’s nonlinear mechanical response~\cite{ewoldt2010large,hyun2011review}.

Amorphous soft matter shows a rich experimental phenomenology in cyclic shear. Early work on athermal suspensions at low volume fractions (i.e., in the liquid rather than solid regime) uncovered a transition from reversibility at low strain amplitudes to irreversibility at high amplitudes, with the timescale to steady state diverging as the transition amplitude was approached from either side~\cite{Pine2005,Corte2008}. An onset of irreversible behaviour with increasing strain amplitude has likewise been reported in amorphous soft solids: in dense emulsions~\cite{hebraud1997yielding} and colloidal glasses~\cite{petekidis2002rearrangements}. The rising incidence of plastic events across the transition was explicitly characterised in Ref.~\cite{keim2013yielding}.  A  sharp peak in the system's correlation time occurs at the transition~\cite{knowlton2014microscopic}, along with spatially growing clusters of plastic events~\cite{HimaNagamanasa2014}.  

In cyclic shear stress, fatigue followed by failure after a delay time that increases with decreasing loading amplitude was reported in a protein gel~\cite{saint2017predicting}, carbon black~\cite{perge2014time,gibaud2016multiple} and hot asphalt~\cite{Kun2007}.  Detailed observations showed the process of fatigue and failure to comprise four successive regimes as a function of increasing cycle number: an initially solid response, followed by plug flow with wall slip, then fluid-solid coexistence, and finally full fluidisation~\cite{gibaud2016multiple}. In this way, strain localisation is seen to  be an important feature in the yielding dynamics of a soft material subject to cyclic loading.

Strain localisation in the form of shear banding is likewise associated with fatigue failure in  hard amorphous materials. Crack growth during fatigue has been characterised experimentally in metallic glasses~\cite{ritchie2000mechanics}. Shear band initiation was evidenced in molecular simulations of metallic glasses~\cite{Sha2015}. As in soft materials, the number of cycles to failure increases dramatically with decreasing cyclic loading amplitude~\cite{gilbert1999mechanisms,menzel2006stress}. 

Alongside these experimental observations, particle simulations have played an important role~\cite{Fiocco2013,BhaumikFoffi2021b,Bhaumik2021,yeh2020glass,maity2026fatigue,Priezjev2013,Leishangthem2017,Parmar2019,Das2020,Regev2013,Sha2015,Kawasaki2016,priezjev2023fatigue,Bhowmik2022}. In cyclic loading, delayed failure was observed after a number of cycles that increases with decreasing load amplitude~\cite{Bhowmik2022}, as seen experimentally. In cyclic shear strain, a sharp increase in diffusivity is seen above a critical strain amplitude~\cite{Kawasaki2016,Fiocco2013},  with a dramatic increase in the number of cycles to attain an absorbing state below this critical amplitude~\cite{Kawasaki2016,Fiocco2013}, or to approach failure above it~\cite{priezjev2023fatigue}. The transition displays significant dependence on a suspension's volume fraction relative to jamming/unjamming~\cite{Das2020,Nagasawa2019}, and on the level of sample annealing prior to shear~\cite{BhaumikFoffi2021b,Bhaumik2021,yeh2020glass,maity2026fatigue}.  Clusters of  mobile particles were reported in Refs.~\cite{Priezjev2013,BhaumikFoffi2021b}. However, the size of plastic avalanches shows no apparent signature of the approaching yielding transition, displaying instead a sudden onset at the transition itself~\cite{Leishangthem2017}. Strain localisation and shear banding have been reported during yielding~\cite{Parmar2019}. A transition between absorbing and diffusing states was captured in a minimal particle simulation,  for volume fractions both below and above jamming, with a diverging timescale at the transition~\cite{Ness2020}. 

This rich phenomenology has motivated significant efforts to understand the cyclic shear of amorphous materials theoretically. In this context, lattice elastoplastic models play an important role~\cite{Nicolas2018}. These minimally distill the salient physical ingredients of an amorphous solid by notionally subdividing it into mesoscopic regions, each represented by an elastoplastic element that loads elastically with shear, in between intermittent local plastic yielding events. The stress relaxed in any such event is propagated elastically to surrounding elements in re-establishing force balance. An elastoplastic model was shown to capture yielding above a critical amplitude of cyclic shear and elastic state absorption at lower amplitudes, with a dramatically increasing timescale near the transition, strain localisation above the transition, and a strong dependence on the level of sample annealing prior to shear~\cite{liu2022fate}. Such behaviour has also been reported recently in a similar elastoplastic model~\cite{khandare2026elastoplastic}.  The role of accumulating strain heterogeneity during fatigue, leading to shear banding upon failure, was explored in an elastoplastic description with simplified one-dimensional stress propagation in Ref.~\cite{cochran2024slow}.

Despite the minimal simplicity of their physical ingredients, and their success in capturing the physical phenomenology just described, elastoplastic models have to date been assumed prohibitive to analytical progress, due to the complexity stemming from long-ranged stress propagation across the lattice, significantly limiting the extent to which elastoplastic models can afford a truly mechanistic understanding of cyclic shear. Even simpler models have therefore been devised. These include single state Ehrenfest models~\cite{sastry2021models,mungan2021metastability}, which consider mesoscopic regions of amorphous material as blocks with transition rules between them, but otherwise without coupling. While such models show qualitative agreement with some aspects of experimental behaviour, they do not capture a true yielding transition. More sophisticated versions~\cite{Parley2022,sarkar2025coarse}, incorporating a mean field noise term to mimic interactions between elements due to stress propagation,  do capture yielding above a critical strain amplitude. In taking a mean field approach, however, they discard true spatial interactions and cannot capture the emergence of plastic clusters and avalanches, or the strain heterogeneity and shear banding that are central to the behaviour of amorphous materials in cyclic shear~\cite{HimaNagamanasa2014,gibaud2016multiple,ritchie2000mechanics,Priezjev2013,BhaumikFoffi2021b,Leishangthem2017,Parmar2019,Aime2018,khandare2026elastoplastic,cochran2024slow}.

In this work, we perform extensive numerical simulations of a lattice elastoplastic model in cyclic shear, with full long-ranged stress propagation between elements~\cite{Eshelby1957,Picard2004,Nicolas2018}. The model captures all the key  observations  described above from experiments and particle simulations: slow progression to an absorbing elastic state at low imposed strain amplitudes, slow fatigue and delayed material failure at higher amplitudes, a rapid increase of the timescale to attain these states as the transition amplitude is approached from either side, growing plasticity across the transition, strain localisation and shear banding above it, and a strong dependence of the transition on the degree of sample annealing prior to shear.

The key contribution of this work is then to elucidate a mechanistic physical understanding of this rich phenomenology, by combining a thorough interrogation of our numerical simulations with analytical calculations.  We start by showing that the  evolution to an absorbing elastic state maps to a process of single element diffusion on a finite interval with two absorbing boundaries. The relevant diffusing variable is the local elemental strain variable, which takes two jumps stroboscopically in each cycle. Once the value at any lattice site diffuses below its local yielding threshold, the site's element behaves elastically in subsequent cycles, leading to a gradual accumulation of strain values just below threshold. This physical mapping enables us to obtain a closed form analytical expression for the decay rate of the fraction $f(C)$ of elements that yield per cycle, as a function of cycle number $C$ since the inception of shear. We furthermore obtain an analytical expression for the survival function $S(C)$ -- the fraction of samples that have not yet attained an absorbing state, across an ensemble of samples with different random initial conditions. Finally, we map out the cycle number $C^*$ at which $S(C)$ falls below a small threshold, as a function of the relevant control parameters. 

In the regime of weak activity, these predictions show excellent agreement with our numerical simulations across the full range of cycle number $C$, from shortly after the inception of shear, $C=O(1)$ to final absorption at $C^*=O(100-1000)$. At higher strain amplitudes (but still below the yielding transition), a regime of stronger plastic activity in early cycles causes an initial departure from these analytical predictions. However,  a simple exponential fit to this early regime of strong activity, for small cycle numbers $C=O(1-10)$, allows a prediction of the full decay function, out to large cycle numbers and final absorption for $C=O(100-1000)$. Importantly, this points the way to predicting {\em long-time} material behaviour from measurements of {\em early-time} signatures alone.

Finally, we show that the  same slow accumulation of elements just below their local threshold that led eventually to the absorbing elastic states just discussed also accounts for the slow fatigue that brings the system to the brink of failure, for imposed strain amplitudes above the transition. We evidence this via simulations in which the strain amplitude artificially exceeds the actual imposed amplitude $\gamma_0$ in each half cycle, before reverting to the system's actual state at $\pm\gamma_0$ and resuming the next half cycle as normal. During fatigue,  the stress-strain slope $d\sigma/d\gamma$  evolves towards unity at $|\gamma|=\gamma_0^-$. This signifies perfect elasticity {\em within} cycle, consistent with ultimate elastic state absorption in any regime  where failure is avoided. In important contrast, however, the  slope  of the stress-strain curve in the fictitious regime just {\em beyond} cycle evolves towards zero. This brings the system arbitrarily close to  mechanical instability, such that even a small fluctuation  can finally precipitate sudden and catastrophic material failure.

The paper is structured as follows. Sec.~\ref{sec:model} introduces the elastoplastic model that we shall simulate. Our results are presented in Sec.~\ref{sec:results}, starting in Sec.~\ref{sec:basic} with a discussion of the model's basic phenomenology, which shows a slow progression to an absorbing elastic state for imposed strain amplitudes $\gammai<\gammac$, and slow fatigue followed by material failure for $\gammai>\gammac$. We then elucidate a mechanistic understanding of each regime in turn: of delayed absorption in Sec.~\ref{sec:absorption} and delayed material failure in Sec.~\ref{sec:yielding}. Our conclusions are set out in Sec.~\ref{sec:conclusions}.

\section{Model}
\label{sec:model}

We study an athermal amorphous material subject to quasistatic cyclic shear. To do so, we simulate an array of elastoplastic elements, one on each of $N\times N$  sites of a 2D square lattice~\cite{PicardAjdari2005,Nicolas2018}. The $ij$-th element has a local elastic shear strain $l_{ij}$ and shear stress $\mu l_{ij}$, with elastic constant $\mu$ the same for all elements. We use a scalar approach, discarding all other components of the stress tensor. The macroscopic shear stress is
\begin{equation}
\Sigma=\frac{\mu}{N^2}\sum_{i=1}^{N}\sum_{j=1}^N l_{ij}.    
\end{equation}

In between intermittent local plastic yielding events, each element loads affinely and elastically with the globally applied strain $\gamma$, with $\Delta l_{ij}=\Delta \gamma \;\forall\; i,j$ in any applied strain increment $\Delta\gamma$. When any element's strain energy $\tfrac{1}{2}\mu l^2$ exceeds a threshold $E$,  the same for all elements, the element immediately yields plastically and selects a new strain $l$ from a Gaussian distribution of zero mean and width $\lw$. Force
balance is then re-imposed across the lattice using the 2D Eshelby stress propagator~\cite{Eshelby1957}:  the propagated change to the elastic displacement field $\vecv{u}$ from a  plastic stress change $\tens{S}$ at the origin is given at wavevector $\vecv{k}$ in Fourier space by~\cite{Picard2004} 
\[
\mathbf u=\frac{1}{\mu k^2}
        \left(
            \mathbf I-\frac{\mathbf k\otimes\mathbf k}{k^2}
        \right)
        \cdot
        \left(
            i\,\mathbf S\cdot\mathbf k
        \right).
\]
The corresponding strain and stress changes are
\[
\boldsymbol{\epsilon}
=
        \frac12
        \left(
            \nabla\mathbf u
            +
            (\nabla\mathbf u)^{T}
        \right),
        \qquad
        \boldsymbol{\tau}
        =
        2\mu\boldsymbol{\epsilon},
    \]
of which we consider only the shear component. This decays as $1/r^d$ with distance $r$ from the origin site of yielding, in $d$  dimensions. For the strain-controlled protocol considered here, the zero wavevector mode of the shear stress propagator equals $-1$.

Prior to shear, each element's local strain $l$ is chosen randomly from a Gaussian distribution of zero mean and standard deviation $l_0$.  Small values of  $l_0$ correspond to well annealed samples with high initial order, and large values to poorly annealed, initially disordered samples. Force balance is then imposed across the lattice using the Eshelby propagator.  This reduces the standard deviation of the distribution of local strains by a factor $r$, equal to the square root of the sum
over lattice sites of the square of the Eshelby propagator. We find numerically that $r=0.672$, for large system size $N$.

To simulate cyclic shear of amplitude $\gamma_0$, we query at each numerical step what strain increment $\Delta\gamma$ is needed
to take the least stable element just over its threshold for yielding, with $\Delta\gamma$ positive in the forward part of the cycle and negative in the reverse part. We then increment all elemental strains and the global strain by this amount: $l_{ij}\to l_{ij}+\Delta\gamma$,  and $\gamma\to\gamma+\Delta\gamma$. The element above threshold is then
yielded and force balance reimposed across the lattice. The resulting stress propagation may then cause other elements to exceed their thresholds. These are yielded in turn, and force balance is again reimposed. This avalanching behaviour is repeated iteratively until no elements remain above yield after rebalancing. The avalanche has then run to completion and we proceed to the next strain step.   Once the magnitude of the global strain attains $\gammai$, the direction of shear is reversed. If the magnitude of strain increment $|\Delta\gamma|$ needed to take the least stable element above threshold is less than $0.001$ in any step, according to the procedure just defined, we set $|\Delta\gamma|$ instead to this small value. This avoids taking prohibitively many strain steps for large systems.

We use a stress scale in which the elastic constant  $\mu=1$  and a strain scale in which the yield strain $\ly\equiv\sqrt{2E/\mu}=1$. Parameters to explore are then the imposed strain amplitude $\gammai$, the level of sample annealing prior to shear (recall that small $l_0$ corresponds to highly annealed samples), the width $\lw$ of the Gaussian from which each freshly yielded element selects its post-yield strain, and the system size of our $N\times N$ lattice.

\section{Results}
\label{sec:results}

In presenting our results, we start in Sec.~\ref{sec:basic} by discussing the model's basic phenomenology: slow progression to an absorbing elastic state for imposed strain amplitudes $\gammai<\gammac$, and slow fatigue followed by material failure for $\gammai>\gammac$, with the  critical strain $\gammac$ increasing with increasing sample annealing prior to shear (decreasing $l_0$). We shall then seek to elucidate a mechanistic understanding of each regime in turn: of delayed absorption in Sec.~\ref{sec:absorption} and delayed material failure in Sec.~\ref{sec:yielding}.

\subsection{Absorption vs yielding}
\label{sec:basic}

\begin{figure}[!t]
\includegraphics[width=0.85\columnwidth]{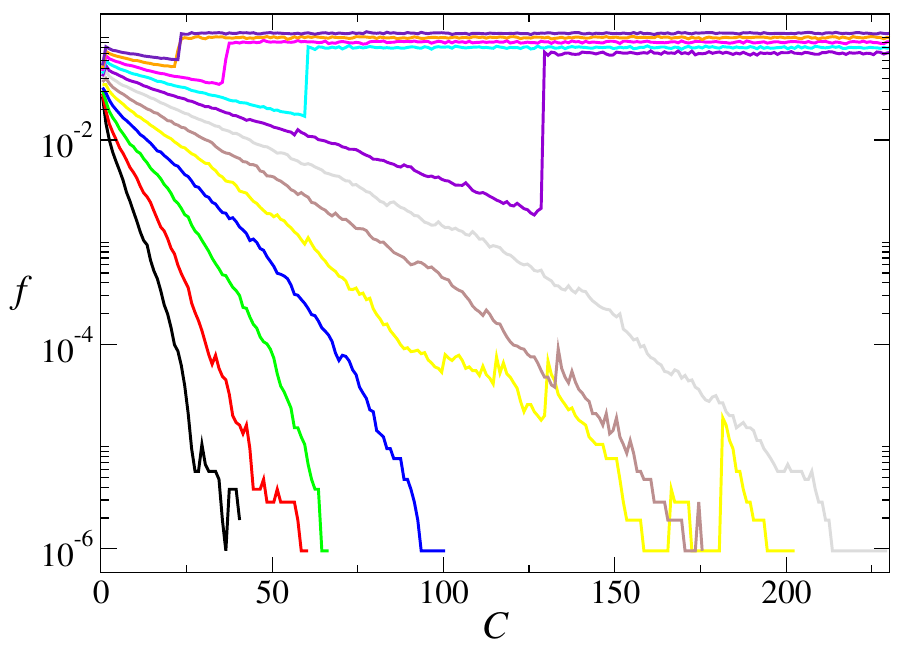}
\caption{Plastic activity vs cycle number for strain amplitudes $\gamma_0=0.730, 0.735\cdots 0.785$, in curves from bottom to top. Each curve represents a single numerical simulation with one value of the random number seed. $l_0=0.22$, $\lw=0.02$, $N=1024$. }
\label{fig:activityVsCycle}
\end{figure}
%scripts/active.sh results/phaseDiagram/cyclicEPM2D2.c_propagator1_gamma00.7{3,4,5,6,7,8}*init0.22*
%

The basic phenomenology of the model is summarised in Fig.~\ref{fig:activityVsCycle}. This shows the fraction $f$ of elements that yield plastically in each cycle  as a function of cycle number $C$ since the inception of shear, for several values of the imposed strain amplitude $\gammai$ at a fixed level of sample annealing $l_0$. We shall call $f$ the ``plastic activity''. For small amplitudes $\gammai<\gammac$, the plastic activity decreases monotonically (to within noise) and falls to zero when the system finally attains an absorbing elastic state.  For larger amplitudes $\gammai>\gammac$, the plastic activity initially decays before suddenly increasing as the sample yields and  fails. The cycle number at absorption (resp. failure) increases dramatically as $\gammai$ approaches $\gammac$ from below (resp. above).

\begin{figure}[!t]
\includegraphics[width=0.99\columnwidth]{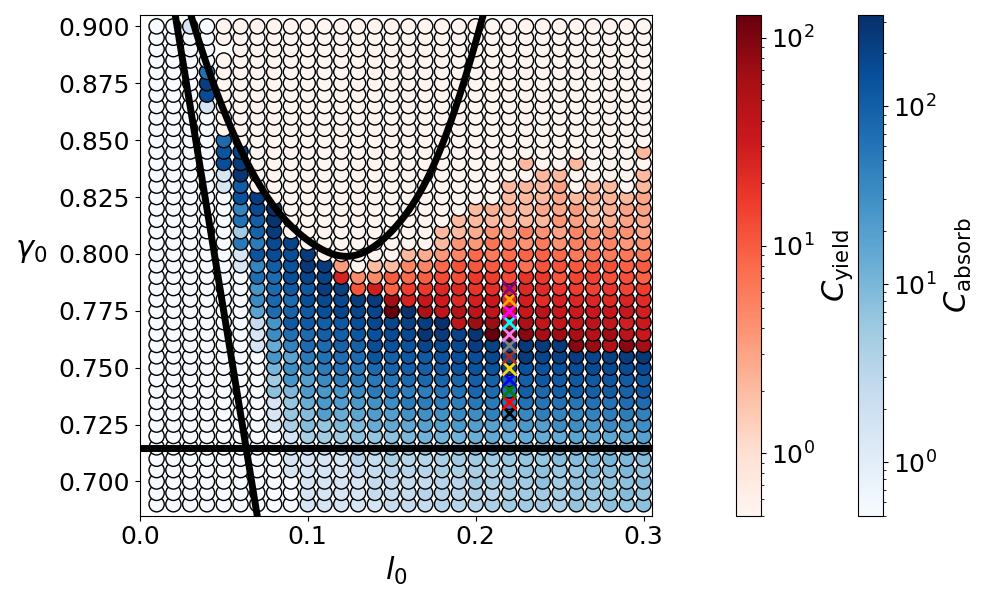}
\caption{Cycle number (averaged across random number seeds) at yielding or absorption in the plane of strain amplitude $\gamma_0$ and annealing parameter $l_0$. Smaller $l_0$ corresponds to better annealed samples.   Blue (resp. red) circles indicate absorption (resp. yielding), with the number of cycles to absorption or yielding shown by the colourbars. The crosses show the locations of the curves in Fig.~\ref{fig:activityVsCycle}. $\lw=0.02$ and $N=1024$. 
}
\label{fig:colourMap}
\end{figure}

The results of Fig.~\ref{fig:activityVsCycle} pertain to a single, relatively poor level of sample annealing prior to shear, as quantified by the parameter $l_0$. Each curve furthermore corresponds to a single numerical simulation with a single value of the random number seed.  The competition between absorption and yielding just described is explored more fully in Fig.~\ref{fig:colourMap}: now in the full plane of $l_0$ and $\gammai$, and now showing  the mean value, across seeds, of the cycle number at absorption or failure. As can be seen, the critical shear amplitude $\gammac$ below (resp. above) which absorption (resp. yielding) occurs increases with increasing  sample annealing prior to shear (decreasing $l_0$). We shall return below to discuss the black lines in Fig.~\ref{fig:colourMap}.

\begin{figure}[!b]
\includegraphics[width=1.0\columnwidth]{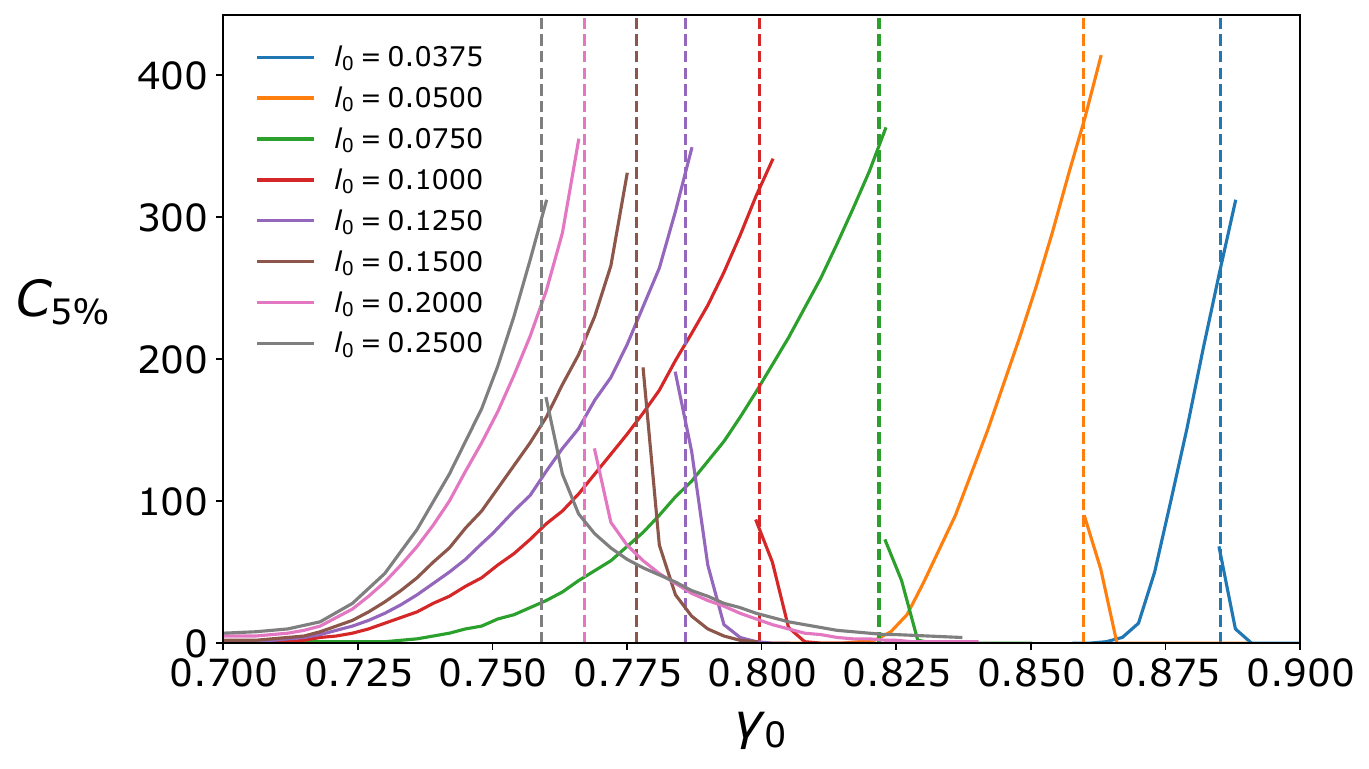}
\caption{Cycle number $\CstTop$ at absorption or yielding versus the imposed strain amplitude $\gamma_0$ for several values of the annealing parameter $l_0$. We define $\CstTop$ as the cycle number at which the survival probability $S(C)$ first falls below $0.05$. Vertical dashed lines show the value of $\gammai$, for each $l_0$, at which the fraction of samples that proceed to an absorbing state and the fraction that fail are equal. $\lw=0.02, N=1024$.
}
\label{fig:Cstar}
\end{figure}

For the poorly annealed system explored in Fig.~\ref{fig:activityVsCycle},  the window of imposed shear amplitudes $\gammai$ over which we see delayed absorption (dark blue symbols in Fig.~\ref{fig:colourMap}) or delayed yielding (dark red symbols)  is relatively broad. For well annealed samples (small $l_0$) this region becomes much narrower, with the region of delayed yielding falling below the discretisation of $\gammai$ in Fig.~\ref{fig:colourMap}.  Delayed yielding does nonetheless arise even for well annealed samples. Fig.~\ref{fig:Cstar} shows the cycle number at absorption or yielding on a finer scale of $\gammai$, for several values of the annealing parameter $l_0$. A clear interval of delayed yielding is evident, even for $l_0$ as low as $0.0375$. (Whereas Fig.~\ref{fig:colourMap} shows the mean cycle to yielding, Fig.~\ref{fig:Cstar} shows the cycle number at which the survival function $S(C)$, defined in Sec.~\ref{sec:absorption} below, falls below $5\%$.) For values of $\gammai$ in the vicinity of the transition, for any $l_0$, some samples fail and some proceed to an absorbing state. The value of $\gammai$ at which these occur in equal ratio is shown as a dashed line for each value of $l_0$.

Having observed this basic competition between delayed absorption and delayed material failure, we now seek to elucidate a mechanistic understanding of both regimes. We start in Sec.~\ref{sec:absorption} with  delayed absorption, considering separately in Secs.~\ref{sec:weak} and~\ref{sec:strong} respectively the regimes in which a small or large fraction of elements are initially active.

\subsection{Delayed absorption}
\label{sec:absorption}

\subsubsection{Weakly active regime}
\label{sec:weak}

\begin{figure}[!t]
\includegraphics[width=0.99\columnwidth]{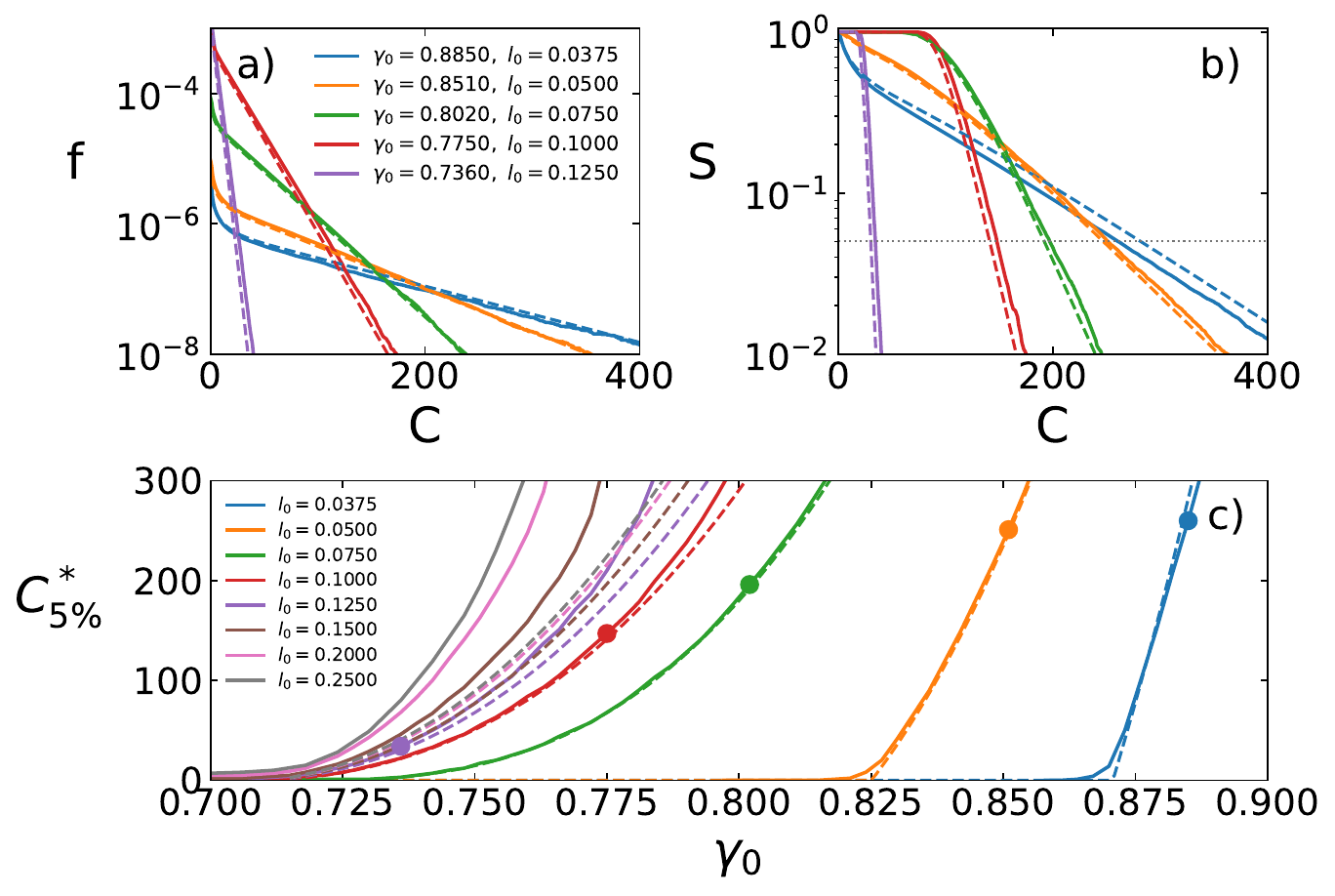}
\caption{Absorbing regime. a) Fraction $f$ of active elements versus cycle number $C$. Data for $f$ is mean averaged over runs performed with different values of the random number seed. b) Survival function: probability of there being at least one active element vs cycle number. (c) Cycle number $\CstTop$ at which the survival distribution in b) intersects the dashed line at $S^*=0.05$, plotted versus the imposed strain amplitude $\gamma_0$ for several values of the annealing parameter $l_0$. Solid lines: simulations of the full elastoplastic model. Dashed lines: analytical predictions of Eqn.~\ref{eqn:dilute_activity} (panel a), Eqn.~\ref{eqn:dilute_survival} (panel b) and Eqn.~\ref{eqn:dilute_Cstar} (panel c). Circles in c) show the location of the runs in a,b). $\lw=0.02$, $N=1024$.
}
\label{fig:dilute1}
\end{figure}

In this section, we aim to understand the system's evolution to an absorbing state, focusing in particular on the regime of low enough strain amplitude $\gammai$ (at any value of the annealing parameter $l_0$), for which only a small fraction of elements is ever plastically active. Accordingly, the active elements  are  dilute: when any such element yields, any resulting stress propagation $1/r^2$ has a negligible effect on the other active elements.  We can therefore proceed, in this weakly active regime, by assuming that the elements behave independently. 

For several combinations of $\gammai$ and  $l_0$,  we show by solid lines in Fig.~\ref{fig:dilute1}a) the evolution with cycle number $C$ of the fraction $f$ of active elements, averaged across many simulation runs (i.e., across random number seeds). After a rapid initial transient, this  shows exponential decay. 

While these data for $f(C)$ averaged across seeds show clean exponential decay, the actual fraction in any individual run fluctuates about this average as it decays. (Examples can be seen in Fig.~\ref{fig:strong1}, in the context of the regime of strong initial activity.) The cycle number at which the activity finally falls to zero therefore varies from run to run. Sampling over many runs then gives the probabilistic survival function $S(C)$. This is defined as the fraction $S$ of runs at any cycle number $C$ for which there remains at least one active element. Our simulation data for this function are shown by solid lines in Fig.~\ref{fig:dilute1}b), with parameters matched to the corresponding activity functions $f(C)$ in Fig.~\ref{fig:dilute1}a). The value of $C=\CstTop$ at which $S$ first drops below $0.05$ is plotted versus $\gammai$ for several $l_0$ in Fig.~\ref{fig:dilute1}c). (These data therefore repeat some of that in Fig.~\ref{fig:Cstar}.)

We now aim to elucidate this behaviour in terms of the evolution from cycle to cycle of the distribution of elemental strains $p(l,C)$ across the lattice. As described in Sec.~\ref{sec:model}, the elements are initialised prior to shear with local strains drawn randomly from a Gaussian distribution of width $l_0$. After force balance is imposed, the resulting distribution of local strains has standard deviation $\siginit=r l_0$:
\begin{equation}
p(l,C=0) = \frac{1}{\sqrt{2\pi}\siginit} \exp\left(-\frac{l^2}{2\siginit^2}\right).
\end{equation}
Here $r=0.672$ is the reduction factor arising from imposition of force balance, as discussed in Sec.~\ref{sec:model}.

In what follows, we shall consider the evolution from cycle to cycle of the distribution $p(l,C)$, stroboscopically at the point in each cycle when the global strain crosses zero in the positive direction. Any element that is inactive throughout a full cycle behaves elastically, with its value of $l$ changing in perfect tandem with the globally applied strain, and being stroboscopically unaltered from cycle to cycle. Any change in $p(l,C)$  therefore arises via the sub-ensemble of elements that do  yield. The elements that yield in the positive half cycle are those for which $l>1-\gammai$ at the stroboscopic point of interest; elements with smaller $l$ do not attain their threshold $\ly=1$ even at the limit of the strain cycle, $\gamma=\gammai$.
%(Likewise in the negative half cycle, elements that yield are those with $l<-(1-\gammai)$ when the global strain crosses zero in the negative direction.) 
Our task is therefore to understand how this subpopulation  depletes from cycle to cycle and eventually falls to zero. The population with $l<-(1-\gammai)$ when the global strain crosses zero in the negative direction behaves likewise, by symmetry.

Consider, then, an element that yields twice in any cycle: once in the forward half, and again in the reverse half. In the positive half, the element yields from a local strain equal to unity, and selects its new strain $\tilde{l}$ from a Gaussian of zero mean and width $\lw$.  Naively, therefore, one would expect a net change $1-\tilde{l}$. In fact the change is $G_0(1-\tilde{l})$ with $G_0=0.571$ for large system size: the Eshelby kernel that describes spatial stress propagation also has an effect at the origin lattice site of yielding, resulting in this modulating factor $G_0$.  In the negative half of the strain cycle, the element yields from a local strain of $l=-1$ and again selects its new value from a Gaussian of zero mean and width $\lw$, with the resulting change again modulated by the factor $G_0$. The two changes $\pm G_0$ cancel. In net effect, therefore, the element  takes two steps of size $\lw G_0\Delta$, with $\Delta$ a random Gaussian variable of zero mean and standard deviation unity.  The local strains of persistently active elements accordingly display single element diffusion with an effective typical step size $\sqrt{2}G_0\lw$ per cycle.

Consider now an element with (stroboscopic) $l$ only just above $1-\gammai$ at cycle $C$. At cycle $C+1$,  having taken two random steps in $l$ value during cycle $C$ as just described, it may still have $l>1-\gammai$ and remain active. Alternatively, it may have stepped across the boundary into the region with $l<1-\gammai$, and thereby become inactive.  Once inactive, it remains so for all future cycles due to the absence of cooperativity between elements in this dilute regime. The strain-diffusion of active elements accordingly has an absorbing left boundary at $l=1-\gammai$. 

Less obvious is the existence of an absorbing right boundary at (stroboscopic) $l=G_0-(1-\gammai)$. (We assume for now that $G_0-(1-\gammai) > 1-\gammai$ and return to discuss the opposite case below.) To understand this, consider an element with a stroboscopic $l>G_0-(1-\gammai)$. In yielding during the positive half cycle, this element's strain reduces by $G_0$, to within a small correction $O(\lw)$. When the global strain subsequently crosses zero in the negative half cycle, therefore, that element's strain satisfies $l>-(1-\gammai)$. Accordingly, by an argument analogous to that for positive strains in the previous paragraph, the element cannot yield in the negative half cycle: it has become inactive, again permanently.

The process of evolution to an absorbing state in this regime of weak activity therefore maps to that of single element diffusion of the local strain values, with step size $\sqrt{2}\lw G_0$ on the interval $[1-\gammai, G_0-(1-\gammai)]$ of width $L=G_0-2(1-\gammai)$, with absorbing boundaries to left and right. The resulting evolution of the distribution of elemental strains $p(l,C)$ is shown in Fig.~\ref{fig:distribution} below. (That figure is discussed below in the context of fatigue before yielding. Before yielding, the evolution of the distribution is essentially the same as in this absorbing regime.)

In the limit of small step size, $\sqrt{2}G_0\lw \ll L$, this process can be solved analytically by projecting the initial distribution $p(l,C=0)$ onto sine basis functions:
\begin{equation}
\label{eqn:A1}
    A_n =  \frac{2}{L} \int_0^L p(\tilde{l}+1-\gammai,C=0) \sin\left( \frac{n \pi \tilde{l} }{L} \right) d\tilde{l},
\end{equation}
with $\tilde{l}=l-(1-\gammai)$. The decay rate per cycle for mode $n$ is $\lambda_n = n^2 \lambda_1$ in which
\begin{equation}
\label{eqn:lambda1}
    \lambda_1 = \frac{1}{2} \left( \frac{\sqrt{2}\lw G_0\pi}{L} \right)^2. 
\end{equation}
Defining the weight factor 
\begin{equation}
\label{eqn:W1}
    W_n = \frac{L}{n \pi} \left[ \cos(0) - \cos\left(n\pi\right) \right]
\end{equation}
and summing over all odd modes ($W_n=0$ for even modes) then gives the fraction of active elements as a function of cycle number $C$
\begin{equation}
\label{eqn:dilute_activity}
    f(C) = 2\sum_{n=1,3,\dots}A_n \, W_n \, \exp(-\lambda_n \, C).
\end{equation}
The prefactor 2 accounts for there being two wings of the distribution.

To account for a non-zero step size, we must apply a correction to this calculation. To do so, we define the Milne extrapolation length $\delta = 0.583005 \, \sqrt{2}\lw G_0$~\cite{BurschkaTitulaer1981}, the effective domain width $L_{\text{eff}} = L + 2\delta$, and the effective wavenumber $k_n = n\pi/L_{\text{eff}}$.  In place of Eqns.~\ref{eqn:A1},~\ref{eqn:lambda1} and~\ref{eqn:W1}, we then have the corrected expressions
\begin{equation}
    A_n =  \frac{2}{L_{\text{eff}}} \int_0^L p(\tilde{l}+1-\gamma_0,C=0) \sin\left( \frac{n \pi (\tilde{l} + \delta)}{L_{\text{eff}}} \right) d\tilde{l},
\end{equation}
\begin{equation}
    \lambda_1 = \frac{1}{2} \left(  \frac{\sqrt{2}\lw G_0\pi}{L_{\text{eff}}} \right)^2, 
\end{equation}
and
\begin{equation}
\label{eqn:Milne}
    W_n = \frac{L_{\text{eff}}}{n \pi} \left[ \cos(k_n \delta) - \cos\left(k_n (L + \delta)\right) \right].
\end{equation}
Using these in Eqn.~\ref{eqn:dilute_activity} gives our final expression for the decay of the plastic activity $f(C)$ in this dilute regime. This prediction is shown by dashed lines in Fig.~\ref{fig:dilute1}a), and agrees well with our full numerical simulations of the lattice elastoplastic model (solid lines).

For an ensemble of $N^2$ independent elements, the survival function $S(C)$ is given in terms of $f(C)$ as:
\begin{equation}
\label{eqn:dilute_survival}
    S(C) = 1 - \Big( 1 - f(C) \Big)^{N^2}.
\end{equation}
Fig.~\ref{fig:dilute1}b) compares this prediction, when used with Eqn.~\ref{eqn:dilute_activity} for $f(C)$, with the full numerical simulation of the lattice elastoplastic model, again showing good agreement. 

\begin{figure}[!t]
\includegraphics[width=0.99\columnwidth]{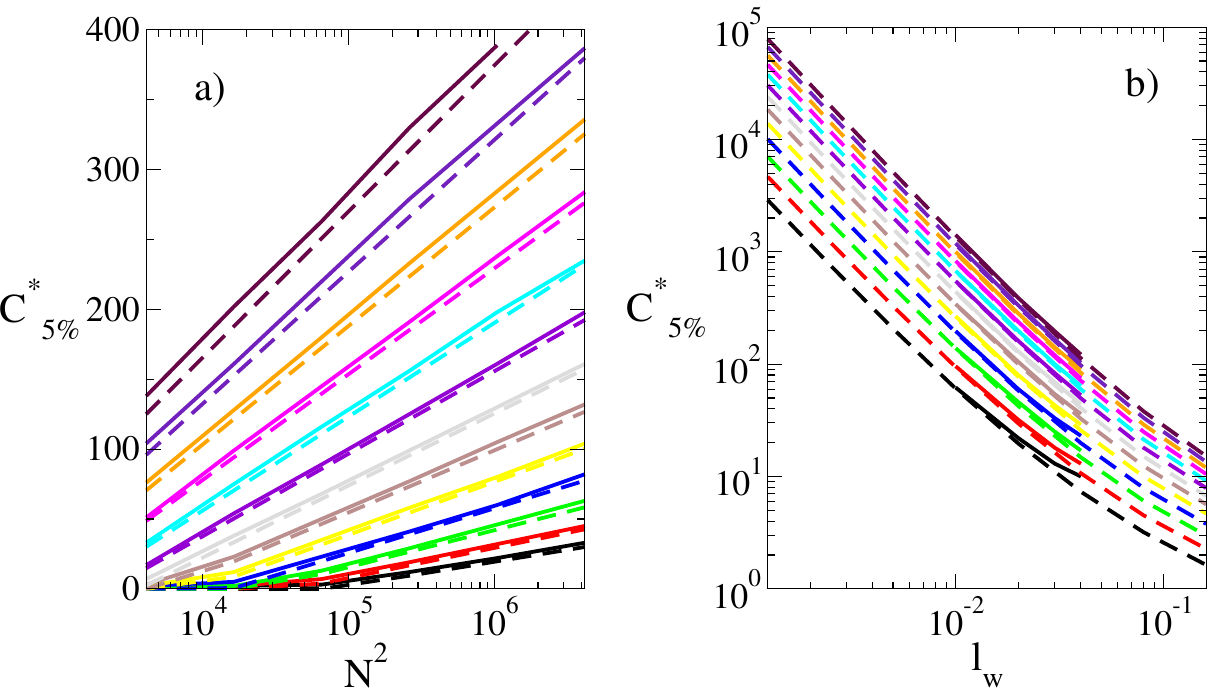}
\caption{Cycle number $\CstTop$ at which the survival distribution intersects $0.05$, plotted versus a) the system size $N^2$ and b) the width $\lw$ of the Gaussian distribution of post-hop local strain values, for values of the imposed strain amplitude $\gammai=0.754, 0.760, 0.766,0.772\cdots 0.826$ in curves upwards in each panel. Solid lines: simulation of the full elastoplastic model. Dashed lines: analytical predictions from Eqn.~\ref{eqn:dilute_Cstar}. $l_0=0.075$. $\lw=0.02$ in a). $N=1024$ in b).}
% solid lines - Sam's data
% survival/top5percent/Cstar_vs_lw_Sam_l00.075_gamma00.{754,76_,766,772,778,784,79_,796,802,808,814,820,826}*N1024*
%survival/top5percent/Cstar_vs_N_Sam_l00.075_lw0.02_gamma00.{754,76,766,772,778,784,79,796,802,808,814,82,826}.dat 
% dashed line - analytical prediction, output using codes integral_vsN.c and integral_vslw.c (they numerically evaluate the projection integral then do the algebra
\label{fig:dilute2}
\end{figure}

We consider finally the cycle number $C^*$ at which the survival function falls below some small threshold $S^*$. Because we are interested in parameter regimes for which absorption is delayed, with large $C^*$, and because modes with $n>1$ in Eqn.~\ref{eqn:dilute_activity} decay over fewer cycles when compared with mode $n=1$, we can retain only that lowest mode. Doing so gives
\begin{equation}
\label{eqn:dilute_Cstar}
C^*=\frac{1}{\lambda_1}\log\left(\frac{2A_1W_1}{1-(1-S^*)^{\frac{1}{N^2}}}\right).
\end{equation}
Fig.~\ref{fig:dilute1}c) compares this prediction to our results for $C^*$ from the full numerical simulation, plotted as a function of $\gammai$ for several $l_0$, showing good agreement for well annealed samples, and for poorly annealed samples at low imposed strain amplitudes. We return below to discuss the departure for poorly annealed samples for larger imposed strains. Fig.~\ref{fig:dilute2} plots $C^*$ as a function of system size $N$ (panel a) and diffusive step size $\lw$ (panel b).

For small $S^*$ (large $C^*$) and large system size $N$, Eqn.~\ref{eqn:dilute_Cstar} approximates to
\begin{equation}
\label{eqn:dilute_Cstar_approx}
C^*\approx \frac{1}{\lambda_1}\log\left(\frac{2A_1W_1 N^2}{S^*}\right),
\end{equation}
which diverges for large system sizes  $N\to \infty$ as
\begin{equation}
\label{eqn:dilute_Cstar_diverge}
C^*\sim \frac{1}{\lambda_1}\log N \sim \frac{L_{\rm eff}^2}{\lw^2}\log N,  
\end{equation}
to leading order, consistent   with the scalings with $N$ and $\lw$ seen in Fig.~\ref{fig:dilute2}. 

It is worth noting that this expression, which holds in the regime of slow absorption (large $C^*$), depends only on the imposed shear amplitude $\gammai$ (via $L_{\rm eff}=G_0-2(1-\gammai)+2\delta$), the width $\lw$ of the post-hop Gaussian, and the system size $N$. It does not depend on the width of the initial Gaussian $l_0$ and is accordingly insensitive to the level of sample annealing prior to shear.  The logarithmic dependence on system size reflects the extreme value statistics of absorption, which completes only when the last surviving element is below threshold. 

These arguments concerning single-element diffusion rest on the existence of an interval $[1-\gammai, G_0-(1-\gammai)]$ of width $L=G_0-2(1-\gammai)$ along the $l-$axis. For imposed strain amplitude $\gammai < 1-G_0/2=0.714$, $L<0$, and the arguments no longer hold. In this regime, below the horizontal black line in Fig.~\ref{fig:colourMap}, we see instead rapid progression to an absorbing state in a small number of cycles. 

For well annealed samples with small $l_0$, the initial Gaussian distribution of local strains is narrow. For sufficiently large system size $N$, this distribution will have non-zero weight even in the far tails of the distribution, with (stroboscopic) $l>1-\gammai$, and the process of single-element diffusion will ensue (for $\gammai > 1-G_0/2$). For realistic system sizes, however, the probability of any element existing in these tails is extremely small. The threshold at which this probability falls below a negligible level is shown by the diagonal black line in Fig.~\ref{fig:colourMap}, for the system size $N=1024$. To the left of this line, the probability of any element being active even in the first cycle is negligible and most samples show purely elastic response from the outset.

\subsubsection{Strong initial activity}
\label{sec:strong}

In the previous section, we considered the decay of plastic activity from cycle to cycle in the weakly active regime, for combinations of $l_0$ and $\gammai$ for which only a dilute fraction of elements is ever active. We now turn to the opposite case of strong activity. Here the active elements are sufficiently concentrated, in early cycles at least, to result in significant elastic interaction spatially via  the Eshelby stress propagator. Because the elements no longer behave independently, the single element diffusion calculation of the previous section no longer applies. It is this that causes the departure of our prediction for $\CstTop$ from the numerical results for poorly annealed samples subject to large imposed strain amplitudes in Fig.~\ref{fig:dilute1}c).

\begin{figure}[!t]
\includegraphics[width=0.99\columnwidth]{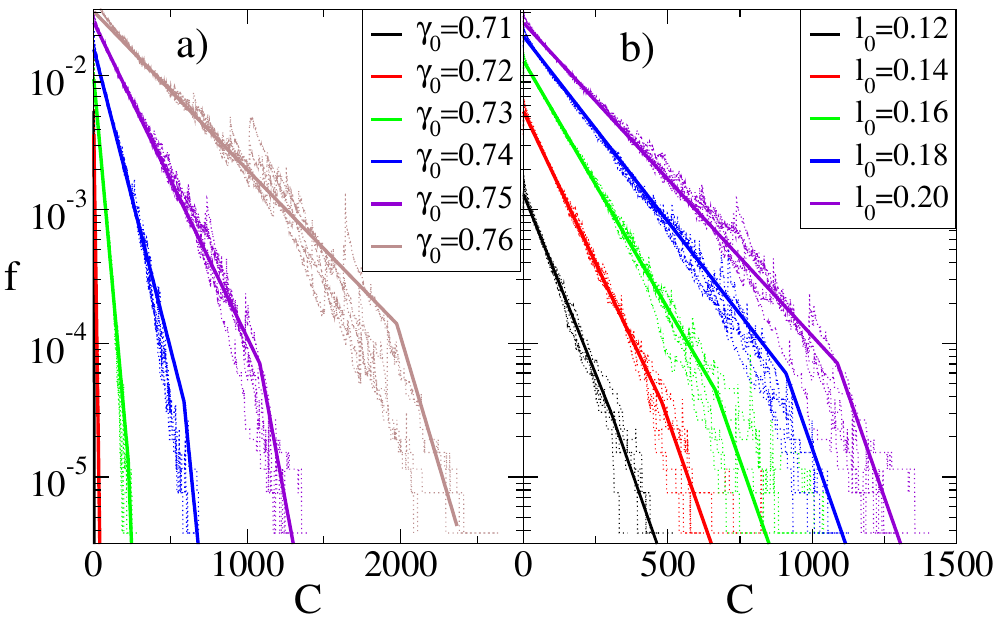}
\caption{Activity versus cycle number for (a) $\gamma_0=0.71\cdots 0.76$ for $l_0=0.20$, (b) $l_0=0.12,0.14\cdots 0.20$ for $\gamma_0=0.75$. Thin dotted lines: simulations of the full elastoplastic model for five different values of the random number seed for each set of parameter values. Solid lines: fits discussed at the end of Sec.~\ref{sec:strong}. $\lw=0.005$, $N=512$.
}
% fits produced by ode.c, files in analytical/ode*dat
\label{fig:strong1}
\end{figure}

Results for the decay of activity $f(C)$ are shown in Fig.~\ref{fig:strong1}: for several values of $\gammai$ at fixed $l_0$ in panel a) and several values of $l_0$ at fixed $\gammai$ in b). For each combination, data are shown by thin lines for five values of the random number seed. Recall that in the weakly active regime,  $f$ showed mono-exponential decay (Fig.~\ref{fig:dilute1}). In the strongly active regime we instead see two exponentials, each with its own decay exponent.  At large cycle number, once the activity has fallen sufficiently that the active elements are again dilute, we recover the results of the weakly active regime, with exponent predicted by the single element calculation of Eqns.~\ref{eqn:dilute_activity} to~\ref{eqn:Milne}.  At smaller $C$, where the activity is stronger, the decay exponent is smaller than that predicted for weak activity. 

\begin{figure}[!t]
\includegraphics[width=0.9\columnwidth]{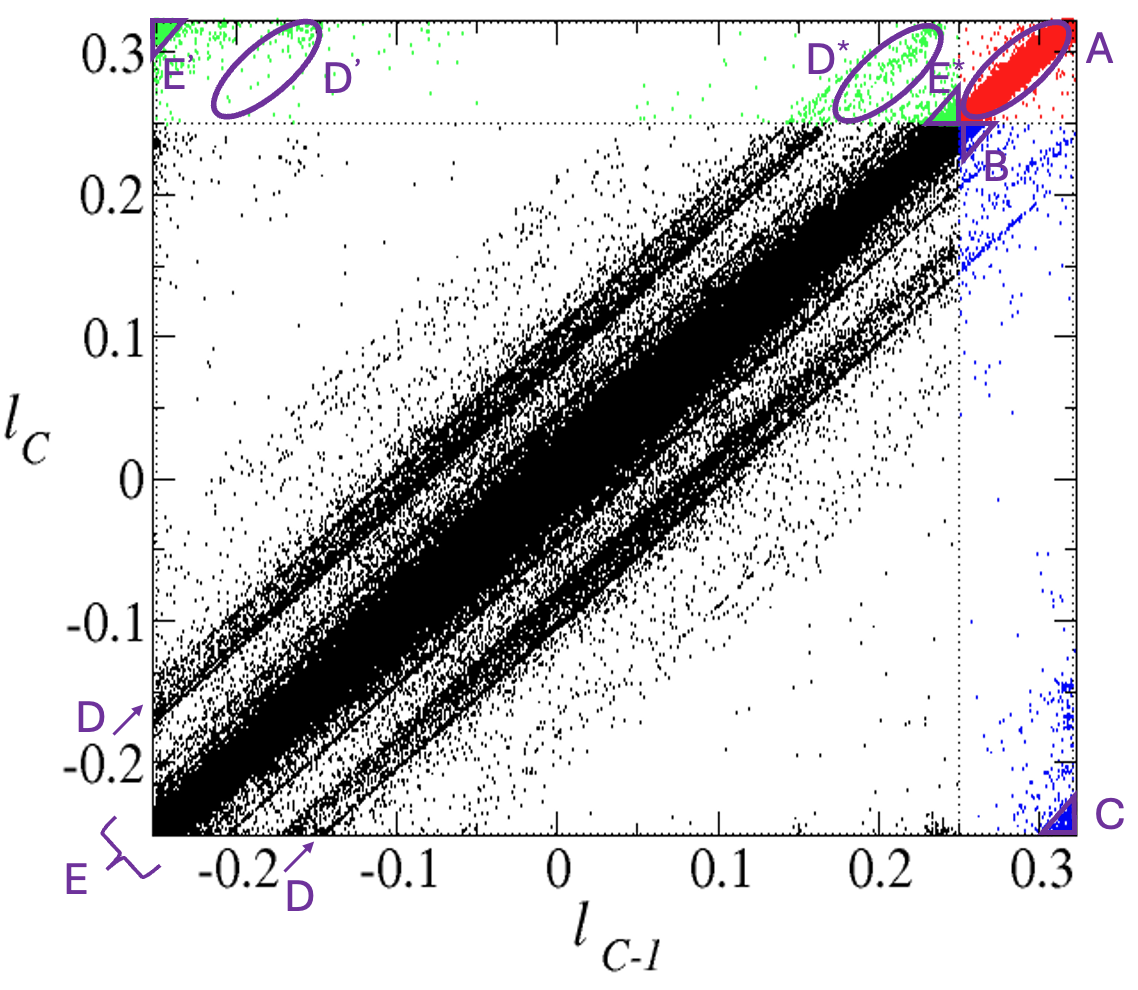}
\caption{Scatter plot of the strain $l$ of each element in cycle $C$ versus that in cycle $C-1$, shown at the point in each cycle where the globally applied strain $\gamma$ passes the origin in the positive direction.  Data are collated for 1000 runs, each with a different value of the random number seed. Plotting all points would lead to prohibitively many points along the line $l_{C}=l_{C-1}$. These points have been filtered out to save memory, with no change to the visual appearance of the plot. Dashed lines indicate the left boundary at $l=1-\gammai=0.25$ and right boundary $l=G_0-(1-\gammai)=0.321$, as discussed in the main text. $\gammai=0.75, l_0=0.18, \lw=0.005, N=512, C=656$.}
\label{fig:strong2}
\end{figure}

To understand this, we consider  again the evolution of the ensemble's elemental strain values from cycle to cycle, stroboscopically at the point in each cycle when the globally applied strain  crosses zero in the positive direction.  Fig.~\ref{fig:strong2} shows a scatter plot of each element's value of $l$ stroboscopically at cycle number $C$ plotted versus that at cycle number $C-1$. Were all elements to behave elastically over the cycle, none would change their stroboscopic $l$, and all dots would lie along the diagonal line $l_C=l_{C-1}$. Any deviations from that line accordingly arise from plastic activity. Several features of this  plot are instructive in helping us to understand this regime of strong activity. 

Dots shown in red correspond to elements that are active at both cycles $C-1$ and $C$, with stroboscopic $1-\gammai<l<G_0-(1-\gammai)$. The strain of each such element takes a step of size $G_0(-1+\lw \xi_1)$ in the first half cycle and of size $G_0(1+\lw \xi_2)$ in the second half. Here $\xi_1$ and $\xi_2$ are independent random variables, each drawn from a Gaussian distribution of zero mean and unit width. The contributions $\pm G_0$ cancel, leaving net hop sizes distributed according to a Gaussian of zero mean and width $\sqrt{2}G_0\lw$. This gives the off-diagonal scatter to the red dots indicated by ellipse A. This small $O(\lw)$ net change to each element's value of $l$ results in only small stress propagation to nearby elements, and is not central to what follows.

Blue dots correspond to elements that were active in cycle $C-1$, with stroboscopic $1-\gammai<l_{C-1}<G_0-(1-\gammai)$, but that become inactive by cycle $C$, with stroboscopic $l_C<1-\gammai$. Consider among these the subset near the point $l_{C-1}=l_C=1-\gammai$, indicated by triangle B. The strain of each such element at cycle $C-1$ just exceeded $1-\gammai$. It then took a step of size $G_0(-1+\lw \xi_1)$ in the first half cycle and of size $G_0(1+\lw \xi_2)$ in the second half, with the net step $O(\lw)$ taking them just below the left absorbing boundary into inactivity, with a strain just below $1-\gammai$. More interesting is the subset indicated by triangle $C$, with stroboscopic $l_{C-1}$ only slightly below $G_0-(1-\gammai)$. These elements took a strain step $G_0(-1+\lw \xi_1)$ in the first half cycle, with the random number $\xi_1$ happening to be such that their $l$ then just exceeds $-(1-\gammai)$ in the second half cycle: these elements cross the right absorbing boundary to become inactive after the first half cycle. Their net strain change over the cycle is therefore $G_0(-1+\lw \xi_1)$, the magnitude of which is much larger than $\lw$. Recall $G_0=0.571=O(1)$.

This $O(1)$ change in local strain of such elements is important, because it results in significant stress propagation to other elements. Their immediate neighbours suffer a stress change of amplitude $0.084$, this being the amplitude of the Eshelby propagator at these neighbouring sites. This causes the stripes marked $D$ in Fig.~\ref{fig:strong2},  offset from the diagonal $l_C=l_{C-1}$ by $0.084$ (with some broadening due to random corrections $O(\lw)$).  Beyond these neighbours, more distant elements also suffer significant stress change via the long ranged Eshelby propagator $~\sim 1/r^2$, with a distribution of Cauchy form. This causes the broad band $E$ about the diagonal.

Some of this stress change indicated by stripes $D$ and band $E$ causes elements that were inactive in cycle $C-1$ to be boosted across the left boundary into subregions $D*$ and $E*$, or across the right  boundary into $D'$ and $E'$. All these subregions reside in the green part of Fig.~\ref{fig:strong2}, containing elements that were inactive in cycle $C-1$ and that have been rendered active by cycle $C$ via stress propagation. It is this effect of elements being (re)activated that causes the decay exponent at early cycles of Fig.~\ref{fig:strong1} to be smaller than that predicted analytically for the regime of  weak activity, in which the left and right boundaries were purely absorbing.

%
%cyclicEPM2D2.c_propagator1_gamma00.75_init0.180_width0.005_log2Nx9_log2Ny9_dgamma0.001_Cmax1000_seed{1..1000}_scatter
%scripts/scatter.sh

It remains an open challenge  to predict the early decay exponent analytically on the basis of such arguments. Here, we instead capture the data in  Fig.~\ref{fig:strong1} by simply fitting (not predicting) the early regime $f_{\rm early}=f_0\exp(-\lambda_{\rm early} C)$. However, we do then make the predictive assumption that once $|df_{\rm early}/dC|<\alpha/ N^2$, with $\alpha=0.1$, negligibly few elements in practice make the important $O(1)$ stress jumps per cycle, on a lattice of size $N\times N$, such that stress propagation is no longer important. At this point we assume a piecewise continuous crossover to the later exponential with dilute dynamics, in which the decay exponent and survival function are predicted via  Eqns.~\ref{eqn:dilute_activity} to~\ref{eqn:Milne} above. So while we fit rather than predict the early regime,  we do predict the crossover, late cycle decay of activity and survival function from knowledge of only the dynamics over the first few cycles. This suggests a possible route to predicting the behaviour of a material after many cycles from its behaviour in early cycles. 

\subsection{Yielding regime}
\label{sec:yielding}

So far, we have studied cyclic shear in the absorbing regime, in which the plastic activity decreases with cycle number and eventually ceases, leaving a purely elastic state. We turn now to the yielding regime. Here   the activity decays initially, before suddenly increasing as the system yields. The cycle number at yielding increases dramatically as the imposed strain amplitude $\gammai$ approaches a critical value $\gammac$ from above, with $\gammac$ increasing with increasing sample annealing (decreasing $l_0$). Recall Figs.~\ref{fig:activityVsCycle} to~\ref{fig:Cstar}. Our aim now is to understand this yielding behaviour, and the process of gradual fatigue that precedes it as the system evolves slowly towards yielding.

\begin{figure}[!t]
\includegraphics[width=0.99\columnwidth]{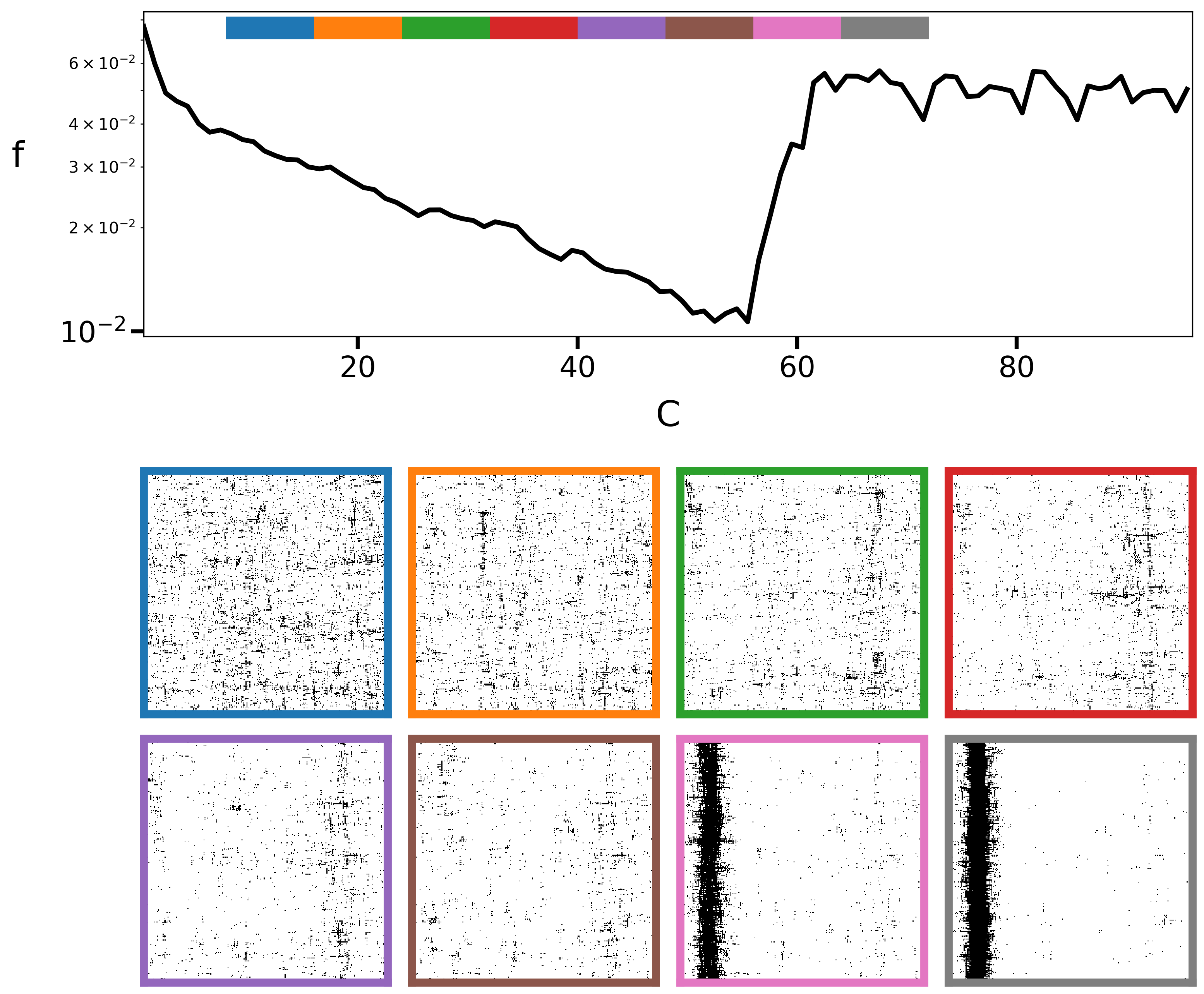}
\caption{{\bf Top:} plastic activity versus cycle number for a single simulation  (i.e., one value of the random number seed). {\bf Bottom:} location, in the $xy$ plane of our simulation lattice,  of plastic events that occur during the intervals of cycle numbers indicated by the coloured bars in the top plot. The colour of each subpanel's frame is matched to the colour of the bar in the top plot. During fatigue, plastic events are relatively few and widely distributed across the system, although with some spatial correlation due to stress propagation. During yielding, the events greatly increase in number and concentrate into a bulk shear band. $\gammai=0.76, l_0=0.3, \lw=0.02, N=256$, as in Fig.~\ref{fig:yielding1}. }
\label{fig:yielding2}
\end{figure}
%cyclicEPM2D2.c_propagator1_gamma00.76_init0.300_width0.02_log2Nx8_log2Ny8_dgamma0.001_Cmax1000_seed1_events
%scripts/events.py
%

We start in Fig.~\ref{fig:yielding2}, the top panel of which shows  the evolution of plastic activity with cycle number. The lower panels show the location of the plastic yielding events in the $x-y$ plane of our simulation lattice, over several intervals of cycle number. In early cycles, the plastic events are scattered widely across the lattice, although with some correlation  due to elastic interactions mediated via the Eshelby propagator. See for example the snapshots framed in blue and orange. This cooperativity increasingly causes the formation of shadowy ``protobands'' of plasticity as the system proceeds towards yielding (e.g., snapshots in purple and brown). A bulk shear band then forms and  the plastic activity  per cycle increases dramatically as the system yields (snapshots in magenta and grey).

\begin{figure}[!t]
\includegraphics[width=0.99\columnwidth]{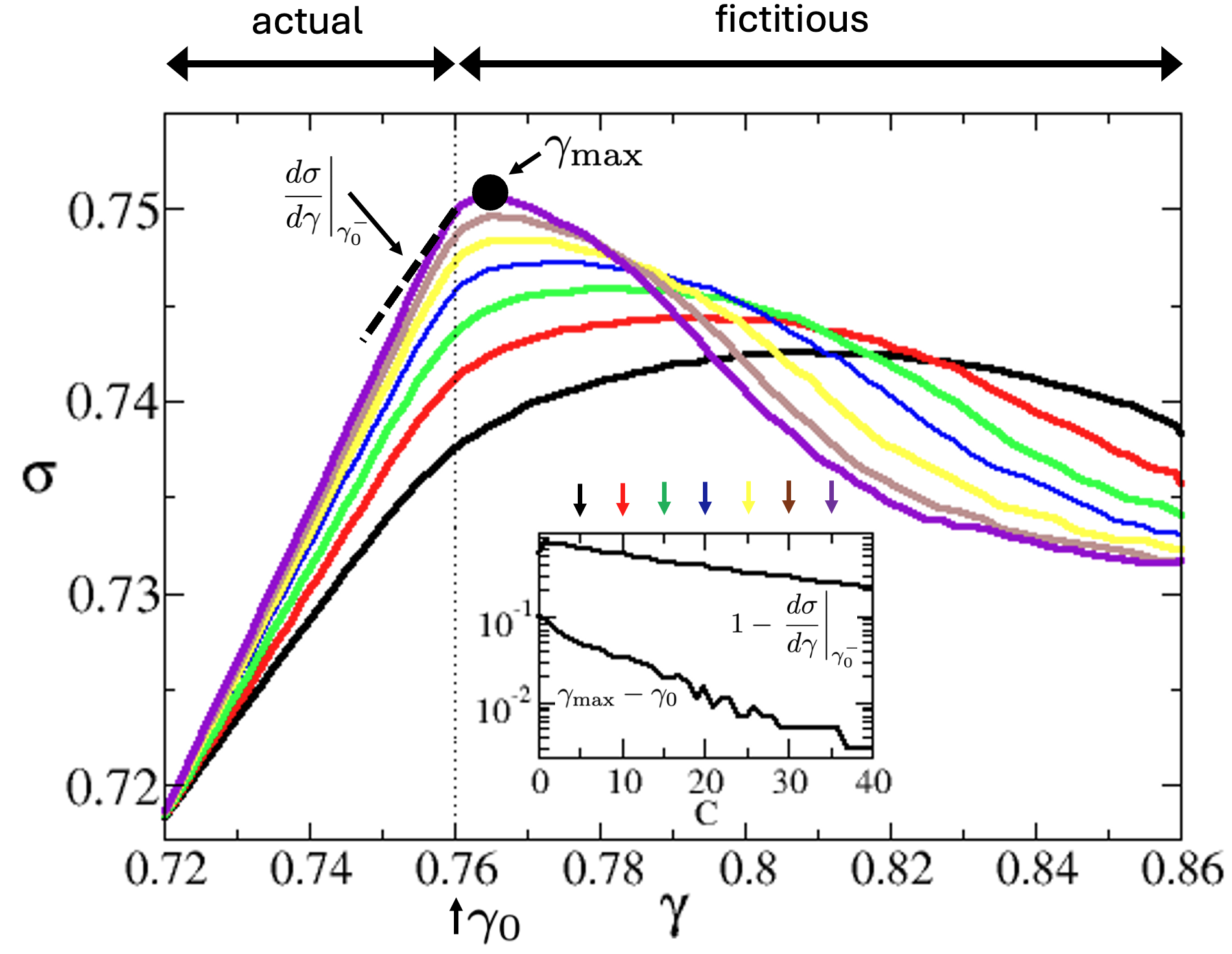}
\caption{Stress-strain curves in simulations in which the strain amplitude artificially exceeds the actual imposed strain amplitude $\gammai=0.76 $ in each half cycle, before reverting to the system's actual state at $\pm\gamma_0$ then resuming the next half cycle as normal. During fatigue, the stress-strain slope $d\sigma/d\gamma$  evolves with increasing cycle number $C$ towards unity at $|\gamma|=\gamma_0^-$, such that $1-\left.d\sigma/d\gamma\right|_{\gamma_0^-}\to 0$ (inset, upper curve), signifying perfect elasticity within-cycle. The location $\gamma=\gamma_{\rm max}$ of the stress-strain maximum in the fictitious regime $\gamma>\gammai$ evolves towards $\gammai$, such that $\gamma_{\rm max}-\gammai\to 0$ (inset, lower curve) and the stress-strain slope $\left.d\sigma/d\gamma\right|_{\gamma_0^+}\to 0$, signifying instability to shear banding just beyond cycle. Once this (fictitious) maximum in $\sigma(\gamma)$ becomes close enough to $\gammai$,  failure via shear banding is triggered by small fluctuations that effectively take the actual state slightly beyond the cycle. The cycle number for each curve in the main panel is indicated by the arrow of corresponding colour just above the inset. Stress-strain curves are averaged over runs for 1000 values of the random number seed. $l_0=0.3$, $\lw=0.02$ and $N=256$.}
\label{fig:yielding1}
\end{figure}

To understand how this arises, we next perform modified simulations in which the strain amplitude artificially exceeds $\gamma_0 $ in each half cycle, before reverting to the system's actual state at $\pm\gamma_0$ and resuming the next half cycle as normal. Fig.~\ref{fig:yielding1} shows the stress versus strain around the upper limit of the positive half cycle, both in its ``actual'' ($\gamma<\gammai$) and ``fictitious'' part ($\gamma>\gammai$), for several values of cycle number $C$.  

Two features of this plot are instructive. The first is the stress-strain slope $d\sigma/d\gamma$  at the limit  $\gamma=\gammai^-$  of the actual cycle. This evolves towards unity with increasing cycle number ($1-d\sigma/d\gamma|_{\gamma=\gammai}$ tends to zero, as seen in the inset),  signifying an evolution towards perfect elasticity within cycle. 
The second is the occurrence of a stress maximum  at a location $\gammamax$ in the fictitious part of the cycle, $\gamma>\gammai$. This location evolves with increasing cycle number towards $\gammai$ ($\gammamax-\gammai$ tends to zero in the inset), such that the stress maximum eventually becomes arbitrarily close to the upper limit of the ``actual''  cycle,  $\gamma=\gammai$. 

A maximum in a stress-strain curve, followed by a region of negative slope $d\sigma/d\gamma<0$, signifies an instability to shear banding.  The insight that these artificially amplitude-extended simulations afford, therefore, is that a shear banding instability develops beyond the cycle and slowly evolves to become arbitrarily close to the upper limit of the actual cycle. Eventually,  even a small fluctuation due to noise or disorder will result in the formation of a shear band, leading to catastrophic material failure.

Whether the system will indeed yield into a shear banded state, or avoid yielding and instead evolve to attain an absorbing elastic state, can therefore be viewed as a basic competition between the rate at which the level of plastic activity that occurs within cycle decays, and the rate at which a shear banding instability that develops beyond the cycle approaches the strain interval of the cycle itself.

\begin{figure}[!t]
\includegraphics[width=0.99\columnwidth]{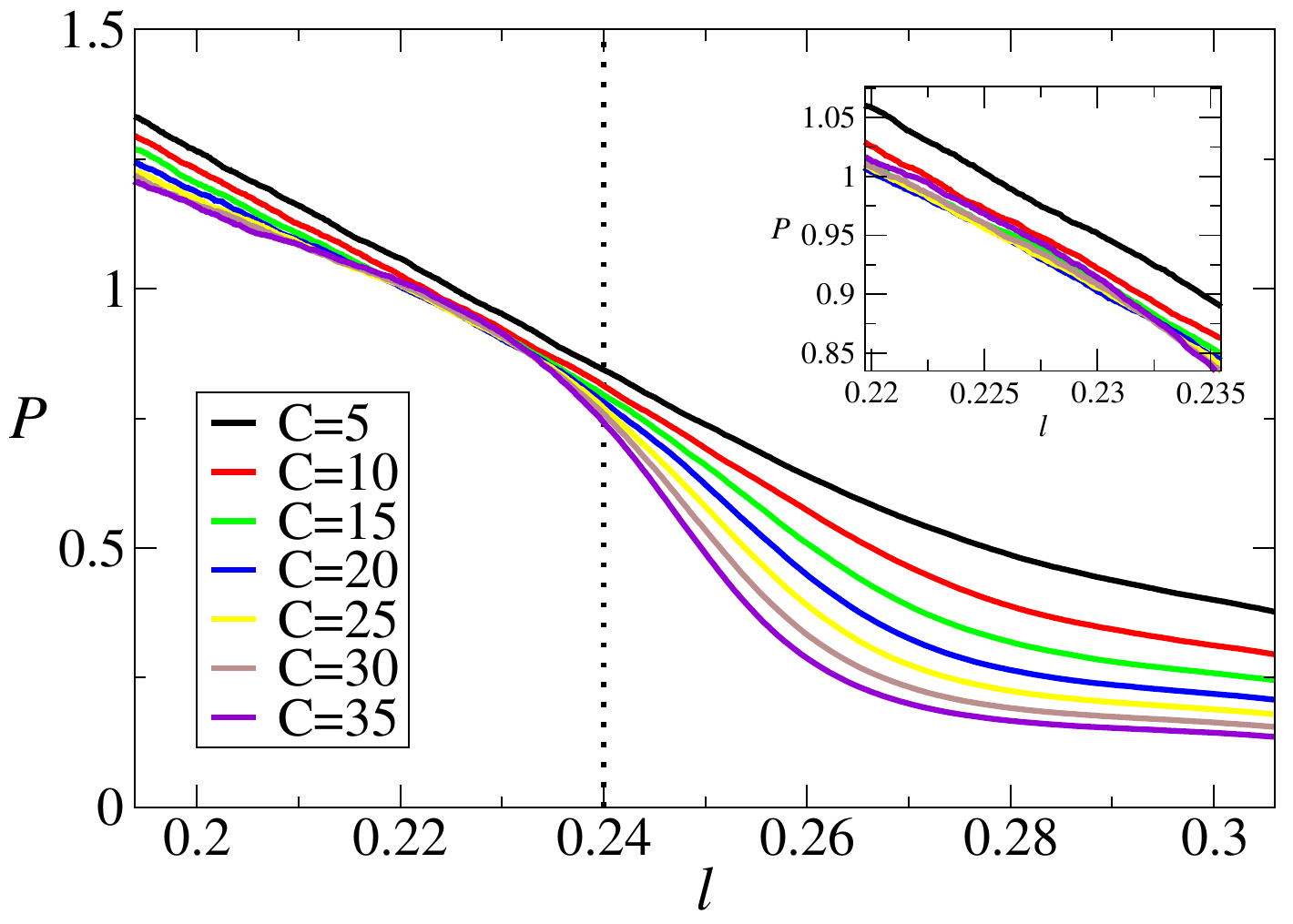}
\caption{Distribution of elemental strains $p(l,C)$ at cycle numbers $C=5, 10, 15\cdots 35$, at the point in the cycle when the globally applied strain $\gamma$ crosses zero in the positive direction. Parameters as in Fig.~\ref{fig:yielding1}. Elements with $l>1-\gammai=0.24$ will cross the yielding threshold $l=1$ in the next quarter cycle, before the global strain attains its maximum $\gammai$. Inset shows a zoom of the interval $l\approx 0.22$ to $l\approx 0.235$.}
\label{fig:distribution}
\end{figure}

To understand the underlying mechanism that governs this competition, we consider finally the evolution of the distribution of elemental strains $p(l,C)$ from cycle to cycle. See Fig.~\ref{fig:distribution}, which shows the distribution at the stroboscopic point in each cycle when the globally applied strain crosses zero in the positive direction. Elements with $l>1-\gammai=0.24$ will be taken across their yielding threshold $l=1$ by the applied shear in the next quarter cycle, before the global strain attains its maximum. As can be seen, this population depletes from cycle to cycle, consistent with a gradual reduction in plasticity within the cycle. Recall the discussion of Fig.~\ref{fig:yielding1}  and Sec.~\ref{sec:absorption} above. Also notable, however, is that a population of elements just below threshold (with $l\approx 0.23$ in Fig.~\ref{fig:distribution}),  slowly accumulates from cycle to cycle. It does so because the active elements from the region $l>1-\gammai$ progressively diffuse across the left absorbing boundary into that inactive regime, as described in Sec.~\ref{sec:absorption}.

It is this accumulation of probability just below $l=1-\gammai$ that leads to the development of a shear banding instability just beyond the cycle. This is because the rate of yielding at any value of the global strain $\gamma$ is set by the probability density $p(l=1-\gamma)$ (at the stroboscopic point of interest), together with the degree to which stress propagation then causes further avalanching. Once the distribution at any value of $l$ becomes large enough, sufficiently many elements would yield, were the globally applied strain increased to $1-l$, to cause a negative stress-strain slope leading to runaway instability and shear banding. 

In this lattice elastoplastic model, therefore, it is this slow accumulation of elements with local strain values just below their yielding threshold, via the diffusion of (previously) active elements across this threshold, that causes slow material fatigue from cycle to cycle and eventually precipitates catastrophic material failure.

We have focused mostly on imposed strain amplitudes  $\gammai$ just above the threshold $\gammac$ that separates yielding from absorption, such that yielding arises  after many cycles of shearing. At larger strain amplitudes, yielding occurs much more quickly. Indeed, this model shows yielding even in unidirectional shear, at a critical strain amplitude that  depends on $l_0$~\cite{Pollard2022}. This amplitude is shown by the upper black line in Fig.~\ref{fig:colourMap}, above which  yielding indeed arises in the first cycle.

\section{Conclusions}
\label{sec:conclusions}

In this work, we have performed comprehensive numerical simulations of a lattice elastoplastic model with full Eshelby stress propagation, and shown it to capture all the key phenomenology exhibited by amorphous materials in cyclic shear. This includes a slow evolution to an absorbing elastic state at low imposed strain amplitudes, versus slow fatigue and delayed material failure at higher imposed amplitudes, with strain localisation and shear banding upon yielding. The timescale to attain each state grows significantly as the transition amplitude is approached from either side, and the transition shows strong dependence on the level of sample annealing prior to the inception of shear.

In the literature to date, elastoplastic models have typically been assumed prohibitive to analytical solution and mechanistic understanding, due to the complexity arising from long-ranged stress propagation across the lattice.  Contrary to this assumption, our key contribution has been to elucidate a mechanistic physical understanding of the phenomena just described, by combining a thorough interrogation of our numerical simulations with analytical calculations. 

We started by demonstrating that the process of evolution to an absorbing elastic state maps to the physics of single particle diffusion on a finite interval with two absorbing boundaries. This enabled us then to obtain analytical predictions for the decay of the fraction of active elements, and for the survival probability of any sample remaining plastically active, as a function of increasing cycle number since the inception of shear. Our analytical predictions show excellent agreement with our numerical simulation results for well annealed samples, and for poorly annealed samples in relatively weak shear.

For poorly annealed samples subject to higher shear amplitudes, strong initial plasticity leads to disagreement between the analytics and numerics at early cycle numbers. The activity later falls to lower levels in which our analytical calculations again apply. A simple exponential fit to the early behaviour over the first few cycles, combined with our analytical results at later cycle numbers, allows a prediction of the material's behaviour at large cycle numbers, including the final progression to an elastic absorbing state. This points the way to potentially predicting a material's behaviour over long timescales, for cycle numbers $C=O(100-1000)$, from measurements of only its early response for $C=O(1-10)$. 

A careful analysis of our simulation data enabled us to elucidate a qualitative understanding of the early regime of strong activity, in terms of the reactivation of elements via Eshelby stress interactions. This physical insight paves the way to future calculations addressing the Cauchy statistics of stress propagation at a mean field level, potentially leading to a quantitative prediction of early time response even with strong activity. 

Considerable attention has been paid in the literature to the development of hyperuniform absorbing states in (initially) disordered systems annealed by shear~\cite{hexner2015hyperuniformity,tjhung2015hyperuniform,weijs2015emergent,wu2015search}. It would be interesting in future work to investigate the existence of such states in elastoplastic modelling approaches, in particular at strain amplitudes just below the transition threshold for poorly annealed samples. While hyperuniformity is typically characterised in a material's concentration field, a search for analogous signals in the strain field appears worthwhile.

In addition to elucidating the dynamics of progression to an absorbing state at low strain amplitudes, our other significant contribution has been to develop a mechanistic understanding of the transition to yielding at larger imposed strain amplitudes. We evidenced in particular that  the accumulation  of elements just below their local yielding threshold during fatigue leads to the development of a shear banding instability for ``fictitious'' strain amplitudes just beyond cycle. The fictitious strain at which that instability occurs decays exponentially with increasing cycle number $C$ towards the actual strain amplitude of the imposed shear.   Any small fluctuation can then precipitate mechanical instability and material failure. It remains an open challenge to elucidate more fully the statistics of how such an instability is triggered.

Our focus in this work has been on amorphous materials subject to cyclic shear of a fixed strain amplitude. In many practical applications, materials are  subject to  a cyclically repeating load. With this motivation, it would be interesting in future to study an elastoplastic model subject to a large amplitude oscillatory shear stress.

{\it Acknowledgments ---}  This project has received funding from the European Research Council (ERC) under the European Union's Horizon 2020 research and innovation programme (grant agreement No. 885146).

%\bibliography{cyclic.bib}


\end{document}